\documentclass[12pt]{article}
\usepackage[cp866]{inputenc}
\usepackage{amssymb}
\usepackage{amsmath}
\usepackage{amsfonts}

\usepackage{graphicx} 
\usepackage{geometry} 

\begin{document}

\author{M.I. Kalinin}
\title{On the status of plane and solid angles in the
International System of Units (SI)}
\date{}
\maketitle

\thispagestyle{empty}%

\begin{abstract}

The article analyzes the arguments that became the basis for declaring in 1995, at 
the 20th General Conference on Weights and Measures that the plane and solid angles 
are dimensionless derived quantities in the International System of Units. The 
inconsistency of these arguments is shown. It is found that a plane angle is not a 
derived quantity in the SI, and its unit, the radian, is not a derived unit. A solid 
angle is the derived quantity of a plane angle, but not a length. Its unit, the 
steradian, is a coherent derived unit of the radian.

\end{abstract}

\section{The 1995 reform of the SI}

In 1960, the 11th General Conference on Weights and Measures
(CGPM), in its Resolution 12 \cite{res12-cgpm11}, adopted the
International System of Units (SI). It included three classes of
units: base units, derived units, and supplementary units. The class of base 
units included the units of length, mass, time, electric current,
thermodynamic temperature, and luminous intensity. The class of derived units
contained 29 units. The class of supplementary units contained a unit of plane 
angle -- the radian, and a unit of solid angle -- the steradian. In 1971, 
the unit of the amount of substance, the mole, was added to the base units.

In the framework of the SI it is considered that the base quantities
have independent dimensions, that is, none of the base units can
be obtained from the others. Derived units are obtained from the base ones 
applying the rules of algebraic multiplication, division and exponentiation. 
Supplementary quantities (the plane and solid angles) also had dimensions independent
of other quantities, and their units were not generated from the base ones.

In 1995, the 20th CGPM adopted Resolution 8 \cite{res8-cgpm20}, which eliminated 
the class of the SI supplementary units, and angles (plane and solid) were 
declared as dimensionless derived quantities. A plane angle was defined as a 
ratio of two quantities having the same dimension of length. A solid angle was 
defined as a ratio of an area to the square of a length. As a result of 
these definitions of angles, their units also became dimensionless. Since 1995 
the unit of plane angle is defined as the dimensionless number ``one'', equal to the 
ratio of a meter to a meter, and the unit of solid angle is also dimensionless
number ``one'', equal to the ratio of a squared meter to a squared meter. The
names radian and steradian can be applied (but not necessarily) for 
the convenience of distinguishing the dimensionless derived units of the 
plane and solid angles. 

What was the reason for such a radical revision made at the 20th CGPM?
And to what extent was it justified? Why were these two dimensional quantities
of different kinds declared dimensionless? One of the principal initiators
of this reform was the International Committee for Weights and Measures 
(CIPM), based on Recommendation U1 of the Consultative Committee for Units (CCU) 
\cite{recom-u1-ccu_80}. CIPM Recommendation 1 of 1980 presented the driving motives \cite{recom1-cipm80}:

\bigskip

\noindent ``{\bf CIPM, 1980: Recommendation 1}
\medskip %

\noindent The Comit\'e International des Poids et Mesures (CIPM),

\medskip %
\noindent {\bf taking into consideration} Resolution 3 adopted by ISO/TC 12 in 1978 
and Recommendation U1 (1980) adopted by the Comit\'e Consultatif des Unit\'es 
at its 7th meeting

\bigskip
\noindent {\bf considering}

\begin{itemize}

\item that the units radian and steradian are usually introduced into 
expressions  for units when there is need for clarification, especially 
in photometry where the steradian plays an important role in distinguishing 
between units corresponding to different quantities, 

\item that in the equations used one generally expresses plane angle as 
the ratio of two lengths and solid angle as the ratio between an area and 
the square of a length, and consequently that these quantities are treated 
as dimensionless quantities, 

\item that the study of the formalisms in use in the scientific field 
shows that none exists which is at the same time coherent and convenient 
and in which the quantities plane angle and solid angle might be considered 
as base quantities,

\end{itemize}

\noindent {\bf considering also}

\begin{itemize}

\item that the interpretation given by the CIPM in 1969 for the class of 
supplementary units introduced in Resolution 12 of the 11th Conf\'erence 
G\'en\'erale des Poids et mesures (CGPM) in 1960 allows the freedom of 
treating the radian and the steradian as SI base units,
 
\item that such a possibility compromises the internal coherence of the SI based 
on only seven base units, 

\end{itemize}

\noindent {\bf decides to} interpret the class of supplementary units 
in the International System as a class of dimensionless derived units 
for which the CGPM allows the freedom of using or not using them 
in expressions for SI derived units.''

\bigskip

Thus, at present, the radian and steradian are defined in the SI 
\cite{SI-8} as follows:
\begin{equation}
\label{rad-sr}%
1\ \mbox{rad} = 1\, \mbox{m/m},  \qquad  1\ \mbox{sr} = 1\,
\mbox{m$^2$/m$^2$}.
\end{equation}                                      
To measure the value of a plane angle, another unit is also used -- a
degree. The degree is not an SI unit, but is allowed to be used on a
par with the radian. And in the SI brochure \cite{SI-8} it is
defined by the expression
$$
1^{\circ} = (\pi/180)\,\mbox{rad}.
$$
It follows that the degree also turns out to be a dimensionless number.

As a result of the 1995 reform, there were numerous  discrepancies, 
inconsistencies, and contradictions in the wordings of the SI.

The first discrepancy concerns the relation between the kinds 
of the plane and solid angles. In the SI, quantities of the same kind have 
identical dimensions, the same units of measurement. They can be compared 
by size, and can be added and subtracted\footnote{This 
circumstance provides an additional effective way to control the correctness 
of mathematical calculations. If the dimensions of individual terms in 
the equation under consideration turn out to be different, then somewhere 
earlier there was an error in the mathematical transformations.}.  
Quantities of different dimensions cannot be compared by size, added 
or subtracted.

The plane angle and the solid angle are qualitatively different quantities. 
The plane angle is a two-dimensional geometric object on a plane, and the solid 
angle is a three-dimensional geometric object in a three-dimensional space. 
They are quantities of different kinds. Comparing the plane angle and 
the solid angle by size is just like comparing the length and the area.
Therefore, the plane angle and the solid angle have different dimensions.

The transfer of the radian and steradian to the class of dimensionless derived 
units resulted in appearing the section ``Units for dimensionless quantities, 
also called quantities of dimension one'' in the SI Brochure \cite{SI-8}. 
The dimensionless quantities are in fact divided into 
three classes there. The first class includes dimensionless 
quantities that are obtained as a result of certain combinations of 
dimensional physical quantities. The unit of such a dimensionless quantity 
is the number one, which is called ``a dimensionless derived unit''. 
It is not clear which base units this number, one, is produced from. 
The Brochure does not specify it.

The second class of dimensionless quantities given in the SI Brochure is the numbers 
that represent counting of objects: a number of molecules, a degree 
of degeneracy of the quantum level, and so on. The unit of these dimensionless 
quantities is also the dimensionless number one. However, this dimensionless 
number one is no longer a derived unit in the SI Brochure, but instead is 
treated as ``a further base unit'' \cite {SI-8}. Somehow, the words 
from Recommendation 1 of the CIPM of 1980 about the need to adhere to a rigid 
scheme with seven base units in SI were forgotten.

And finally, the third class of dimensionless quantities are the plane angle 
and the solid angle mentioned above. The following is said about them: 
``In a few cases, however, a special name is given to the unit one, in order 
to facilitate the identification of the quantity under consideration. This is 
the case with the radian and the steradian. The radian and the steradian have 
been identified by the CGPM as special names for the coherent derived unit one, 
to be used to express values of the plane angle and the solid angle, respectively, 
and are therefore included in Table 3\footnote{Coherent derived 
units with special names and symbols in the SI.}''.

What is this non-dimensional number one, which is marked with two different 
names to distinguish what values it refers to? Indeed there are two
different units of measurement of these quantities after all. Such interpretation of 
this text of the SI Brochure is in complete agreement with the fact that 
the plane and solid angles are quantities of different kinds and are 
measured in different units. There was no need to define them as 
dimensionless quantities, immediately introducing different units of their 
measurement, and without giving a practical definition for these units at that.

The definitions of the radian and the steradian as derived units  
in the SI Brochure \cite{SI-8} also have some internal contradiction. 
According to the definitions given there, the two derived units, the radian 
and the steradian, are expressed in terms of one base unit meter by the 
relations \eqref{rad-sr}. If the usual rules of mathematics are applied in 
these formulas, then the meter is reduced in the numerator and denominator. 
As a consequence, the definitions of angle units will have no base units at all.
And if an equation has no quantity, then nothing in this equation depends on 
the missing quantity.

As a result, the expressions \eqref{rad-sr} will not contain any of the seven 
base units of the SI. Therefore, they cannot be derived from these seven 
base units. That is, the derived units, the radian and the steradian, are  
determined neither through the base unit meter, nor through other base SI units.
And this contradicts the basic concept of the coherent SI system, 
according to which all derived units are determined coherently through 
the base ones. So the assumption that the radian and the steradian \eqref{rad-sr} 
are derived from current base SI units contradicts itself.

Another problem connected with the change in the status of angles in the SI 
adopted in 1995 is the practical realization of the radian and the steradian 
based on their current definitions. It is still impossible to reproduce the radian
and the steradian value in terms of the unit of length using only the definition of \eqref{rad-sr}. An angle of 1 radian, obtained through dividing 1 meter 
by 1 meter cannot be drawn. There are no other definitions of the units of angles 
in the SI brochure.

In practice, the radian is determined using an old well-proved method, by 
constructing the central angle subtended by an arc that is equal in length 
to the radius. Another method is dividing the total plane angle into $2\pi$ 
equal parts, where $\pi$ is the known dimensionless irrational number. 
The unit of the degree is not so difficult to determine, it is an angle 
obtained by dividing the total plane angle into 360 equal parts. And the 
expression of 1 radian in degrees is: 1 rad = 57.2957795$^{\circ}$. 
It is these units which serve to measure of plane angles. But to measure 
an angle with a dimensionless number ``one'' is not possible. A dimensionless 
number is a mathematical concept, an abstraction.

We note here that in the draft of 9th edition of the SI Brochure, prepared
for the 26 CGPM \cite{SI-9}, an attempt was made to take this 
circumstance into account. The definitions of the radian and the steradian in 
that draft (Table 4. The 22 SI units with special 
names and symbols) remained the same as in the 8th edition - dimensionless 
number one. But the explanatory footnotes to these units are very different.

In the 8th edition of the SI Brochure in the footnote ({\it b}), relating to 
the radian and the steradian in Table 3 \cite[P. 118]{SI-8}, the following is 
written:
{\renewcommand{\labelitemi}{({\it b})} %
\begin{itemize}
\item   The radian and steradian are special names for the number one that may be 
used to convey information about the quantity concerned. In practice the symbols 
rad and sr are used where appropriate, but the symbol for the derived unit one 
is generally omitted in specifying the values of dimensionless quantities.
\end{itemize}       }

In the draft of the 9th edition of the SI Brochure the footnotes ({\it b}) 
and ({\it c}), relating to the radian and the steradian in Table 4, \cite[P. 21]{SI-9}
define these units as follows:
{\renewcommand{\labelitemi}{({\it b})}    %
\begin{itemize}
\item   The radian is the coherent unit for plane angle. One radian is the angle 
subtended at the centre of a circle by an arc that is equal in length to the radius. 
It is also the unit for phase angle. For periodic phenomena, the phase angle 
increases by $2\pi$ rad in one period. The radian was formerly an SI supplementary 
unit, but this category was abolished in 1995.

\renewcommand{\labelitemi}{({\it c})}     %
\item   The steradian is the coherent unit for solid angle. One steradian is 
the solid angle subtended at the centre of a sphere by an area of the surface 
that is equal to the squared radius. Like the radian, the steradian was formerly 
an SI supplementary unit.
\end{itemize}          }
These definitions clearly indicate that radians and steradians are not a 
dimensionless number one, but plane and solid angles of certain sizes.

\bigskip

Let us analyze the considerations underlying the 1995 SI reform. 
It can be seen from CIPM 1980 Recommendation 1 text that the main reasons 
for declaring angles as dimensionless derived quantities were:

\begin{enumerate}

\item   The assertion that  plane angle is expressed as the ratio of two lengths
and solid angle is expreesed as the ratio between an area and the square 
of the length.

\item   The assertion that there are no formalisms containing plane and solid 
angles as base quantities, that are at the same time convenient and coherent, 
and that the present SI structure with seven base units is in fact the only 
possible coherent system.

\end{enumerate}

\noindent
Let us consider these assertions in more detail.

\section{Analysis of justifications for transferring angles into
the class of dimensionless derived quantities}

The first assertion is the only basis for transferring plane and solid 
angles into a class of dimensionless quantities. The second assertion serves 
to justify the declaration of both these angles as derived quantities. Let us
consider the above stated assertions more closely.

 In the second statement, it is not clear which scientific formalisms were 
studied and how. Why did the authors of the resolution consider the possibility 
of simultaneously assigning plane and solid angles to either base quantities 
or derived ones? And why is the structure with seven base quantities 
(and their units) considered the only possible? After all, there had been an 
experience of changing the structure of the SI by that time already. In 1971, 
there was a precedent of expanding a list of base units from six to seven units, 
when the amount-of-substance unit was introduced into the SI as the base one, and not 
derived. And this neither caused any inconvenience of work, nor broke the  
coherence of the system of units. In the work \cite{Wittmann}, for example, 
a variant with eight base units is suggested, which in addition 
to the list of seven base units also included the radian.

\begin{figure}  [h]
\center
\includegraphics[width=0.5\linewidth]{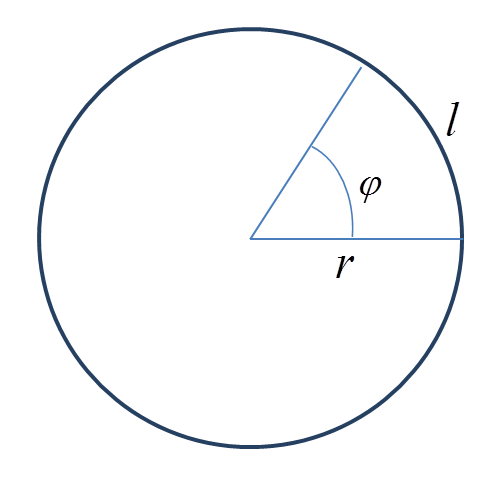}
\caption{In addition to calculating an arc length of radius $r$, 
bounded by an angle $\varphi$}
\label{l-fi-r}
\end{figure}

The first statement is a bit inaccurate and needs to be considered 
more closely. We start with the plane angle. Let us try to examine what the 
formula, connecting an angle and two lengths expresses. To this end, 
we solve the problem of determining the length $l$ of an arc of radius $r$, 
bounded by the central angle $\varphi$. Figure \ref{l-fi-r} shows the arc and 
the corresponding central angle. To solve this problem, the arc is supplemented 
to a circle of the same radius.

It is can easily be seen that the ratio of the length $l$ of the arc to the 
length of the entire circle $2\pi r$ is equal to the ratio of the angle value 
$\varphi$ to the total plane angle value, which we denote by $\Phi$. 
We can write this equation as
\begin{equation}
\label{l-fi-L-Fi}%
\frac{\varphi}{\Phi} = \frac{l}{2\pi r}.
\end{equation}                                      

In metrology, there is a special form of writing any quantity, proposed by Maxwell \cite{Maxwell}, $\varphi = \{\varphi\}[\varphi]$. Here $[\varphi]$ 
is the unit of measurement for $\varphi$, and $\{\varphi\}$ is the numerical value (dimensionless number) of $\varphi$ measured in units of $[\varphi]$. Using this 
form of recording angles in the left-hand side of the equation \eqref{l-fi-L-Fi}, we
can rewrite it as
$$
\frac{\{\varphi\}[\varphi]}{\{\Phi\}[\varphi]} = \frac{l}{2\pi r}.
$$
Here the units $[\varphi]$ in the left-hand side of the equation are simplified 
leaving the ratio of the two dimensionless numbers. Rewriting the resulting equality,
we shall have an expression for $\{\varphi\}$
\begin{equation}
\label{fi/Fi}%
\{\varphi\} = \frac{\{\Phi\}}{2\pi} \cdot \frac{l}{r}.
\end{equation}                                      

Depending on the choice of the unit $[\varphi]$, the ratio will have a different 
form.  If we measure angles in degrees, then $[\varphi] = 1^{\circ}$. 
In this case, the dimensionless number $\{\Phi\}$ is equal to 360, and the formula 
\eqref{fi/Fi} takes the form
\begin{equation}
\label{fi/Fi-grad}%
\{\varphi\} = \frac{180}{\pi} \cdot \frac{l}{r}.
\end{equation}                                      
This coefficient $180/\pi$ (in general $\{\Phi\}/2\pi$) arises in mathematical
 calculations related 
to the angles and functions of them, violating the compactness of mathematical 
formulas. And performing many calculations, it will repeatedly occur making 
them too immense and cumbersome.

Mathematicians have devised a unit of measurement which simplifies 
formulas containing angles. If we choose such a unit of plane angle, which 
when using makes the dimensionless number $\{\Phi\}$ equal to 
$2\pi$, then the expression \eqref{fi/Fi} has this compact form\footnote{Formulas 
of the form \eqref{fi/Fi-grad} and \eqref{fi/Fi-rad} are given in the 
Mathematical Encyclopedia \cite[p.15]{mathencycl-4} in the article "Circle" 
to express the length of the arc through the radius and angle.}
\begin{equation}
\label{fi/Fi-rad}%
\{\varphi\} = \frac{l}{r}.
\end{equation}                                      
The corresponding plane angle unit, ensuring the equality $\{\Phi\}=2\pi$, 
is called the radian (symbol ``rad''). And the radian itself is determined 
based on the condition that the total plane angle is equal to
\begin{equation}
\label{radian}%
\Phi = 2\pi\ \mbox{rad}.
\end{equation}

The expression \eqref{fi/Fi-rad} shows that the ratio of the two lengths  
determines not the quantity angle $\varphi$, but only its numerical value, 
measured in radians. This value $\{\varphi\}$ is indeed a dimensionless 
number by definition. But, contrary to the statement in CIPM 
Recommendation 1 (1980), the expression \eqref{fi/Fi-rad} does not produce
any restrictions on the dimension of the angle itself. As well as no other 
expressions are available, leading to a conclusion that the angles are 
dimensionless. So it is only needed to deduce and employ mathematical 
formulas correctly.

These arguments are also true for the solid angle $\omega$, for which 
it is easy to derive an expression similar to the relation \eqref{fi/Fi} 
for a plane angle
\begin{equation}
\label{omega/Omega}%
\{\omega\} = \frac{\{\Omega\}}{4\pi}\cdot \frac{S}{r^2},
\end{equation}                                      
where $\{\omega\}$ -- the numerical value of the solid angle under 
consideration in units of $[\omega]$, $\{\Omega\}$ -- the numerical 
value of the full solid angle in the same units, $S$ -- the area of 
surface bounded by the solid angle $\omega$ on a sphere of radius $r$ 
centered at the angle vertex.

The unit of solid angle steradian (symbol ``sr'') is chosen, by analogy with 
the unit of plane angle, so that the expression for the numerical
value of the angle $\{\omega\}$ has a compact form of a ratio of 
the area $S$ to the squared radius $r$.
\begin{equation}
\label{omega-s-r}%
\{\omega\} = S/r^2   \qquad \mbox{in the unit steradian}.
\end{equation}                                     
So the ratio of an area to the squared length determines not the solid 
angle  $\omega$ itself, but its numerical value $\{\omega\}$, measured in 
steradians. The steradian can be defined similarly to the radian by setting 
the value of the total solid angle
\begin{equation}
\label{steradian}%
\Omega = 4\pi\ \mbox{sr}.
\end{equation}

Neither physicists nor mathematicians are accustomed to use in mathematical 
calculations with angles the metrological notation of the form $\varphi = \{\varphi\}[\varphi]$. They just write $\varphi$, as in the argument of 
trigonometric functions, where this is really an angle, and in formulas of 
the form $l = \varphi r$, where they actually mean the numerical value 
$\{\varphi\}$ of the angle $\varphi$, measured in radians. This is due to the 
fact that mathematicians work only in a radian measure, which is also typical for 
physicists in their theoretical calculations. And in that case the space of 
plane angle values can be one-to-one connected with the space of real (dimensionless) 
numbers, and all mathematical calculations with angles can be made in the same 
way as with dimensionless numbers. It is really convenient. However, this does 
not mean that angles become dimensionless quantities. If the angle is measured 
in degrees, then both mathematicians and physicists write $\{\varphi\}^{\circ}$, 
if it is measured in grads, then they write $\{\varphi\}$ grad. If the angle 
is measured in radians, then they just write $\{\varphi\}$. Such recording 
means that the angle under consideration is equal to $\{\varphi\}$ radians.

Let us note one more consideration which is related to the effects of the 
general relativity theory (GR). In GR, space is non-Euclidean, it is curved. 
Many formulas of Euclidean geometry become incorrect. In particular, even the 
refined relations \eqref{fi/Fi} and \eqref{omega/Omega} for the dimensionless 
numerical values of the angles are incorrect. In curved space, the length of 
the arc $l(r,\varphi)$ will no longer be a linear function of the radius and 
the numerical value of the angle. But due to the fact that the GR space is 
locally flat, for small values of the radius $r$ this deviation from 
linearity will be small. And the smaller the value of $r$, the more accurate 
are the expressions \eqref{fi/Fi} and \eqref{omega/Omega}. In the limit 
$r\to 0$, these expressions become exact
\begin{equation}
\label{fi-omega-rls}%
\{\varphi\} = \lim_{r\to 0}\frac{\{\Phi\}}{2\pi} \frac{l}{r},
 \qquad \{\omega\} =  \lim_{r\to 0} \frac{\{\Omega\}}{4\pi} \frac{S}{r^2}.
\end{equation}                                      

These expressions also indicate that the angles do not depend on the lengths. 
Moreover, definitions of the radian and the steradian in terms of the arc 
length and surface area in curved space become incorrect for finite values 
of $r$, while the ratios \eqref{radian} and \eqref{steradian} do not depend on 
lengths at all.

In order to determine the status of plane and solid angles in the SI, it is 
necessary to investigate the geometric nature of these quantities and the 
relationship between them.

\section{Analysis of the relationships between plane and solid angles and their units}

\begin{figure}  
\center
\includegraphics[width=0.5\linewidth]{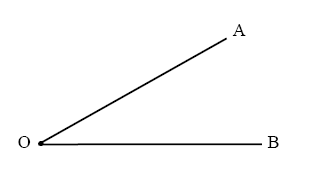}
\caption{Plane angle}
\label{plane}
\end{figure}

First we shall consider the plane angle. It is defined as a geometric figure 
consisting  of two different rays starting from a single point \cite{mathencycl-5}. 
More specifically, the angle represents the entire area of the plane enclosed between 
these two rays. It is usually represented in the form shown in Figure 
\ref{plane}. The rays OA and OB are called the sides of the angle, and 
their common origin O is called the vertex.

By definition, the sides of an angle are not finite segments of straight lines, 
but endless rays. They are depicted as finite segments of arbitrary length. 
The value of a plane angle is determined by the magnitude of the deviation of 
one ray from another when the vertex O is fixed. This deviation does not depend 
on the length of the sides of the angle. They can be increased, or reduced, 
or made of different lengths. The magnitude of deviation of the rays does not 
change in this case. And, therefore, the value of the angle does not change. 
In defining the angle no lengths are involved. Consequently, the value of the 
plane angle does not depend on the length, or any other SI quantities. This 
means that the value of a plane angle is not a derived quantity in the SI.

The feature of the plane angle to characterize the deviation of one ray from 
another is used in mathematics to build a polar coordinate system on a plane, 
as well as cylindrical, spherical, and other kinds of coordinate systems 
in three-dimensional space.

\begin{figure}    [h]
\center
\includegraphics[width=0.5\linewidth]{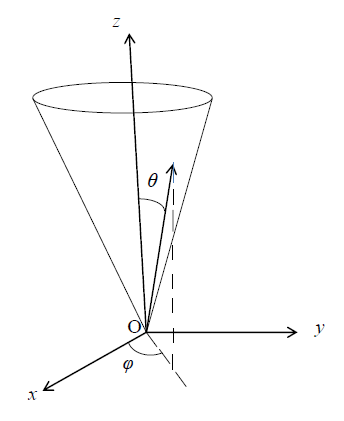}
\caption{Solid angle and spherical coordinate system}
\label{solid}
\end{figure}

Let us now take the solid angle. In \cite{mathencycl-5}, the solid angle is 
defined as the part of space bounded by one cavity of a certain conical 
surface (see Figure \ref{solid}). As in the case with the plane angle, the 
lengths of the rays that make up the conical surface of the solid angle are 
not limited. The spatial direction of these rays is of importance. In contrast 
to the plane angle, the solid angle cannot be defined on a plane. It is a 
three-dimensional object. The conical surface itself is a continuous closed set 
of rays emanating from the vertex of the solid angle.

It is almost obvious that the solid angle is formed from plane angles, like
the area of any two-dimensional region in a plane is formed from straight 
line segments. To show this we construct Cartesian and spherical coordinate 
systems with their common origin at the vertex O of the solid angle, shown 
in Figure \ref{solid}. Any point of three-dimensional space in the chosen 
coordinate system is represented by the vector $\mathbf r$, starting at 
the origin of coordinates and ending at this point. In the Cartesian coordinate 
system, the vector $\mathbf r$ is defined by three coordinates $(x,y,z)$. 
In the spherical coordinate system, the same vector will be defined by 
the coordinates $(r,\theta,\varphi)$, where $r$ is the length of the vector, 
$\theta$ is the plane angle between the $\mathbf r$ vector and the $z$ axis, 
$\varphi$ is the angle between the projected vector on the $(x,y)$ plane and 
the $x$ axis. The range of variables of the spherical coordinate system is 
defined by the expressions: $0\le r<\infty$, $0\le\varphi<\Phi$, 
$0\le\theta \le\Phi/2$. Here we again use the notation $\Phi$ for the value 
of a full plane angle.

The direction of any ray (if we just ignore its length) starting from 
the origin of coordinates is defined in the spherical coordinate system by two 
plane angles $\theta$ and $\varphi$, as shown in Figure \ref{solid}. For each 
value of the plane angle $\varphi$, the corresponding ray of the conical surface 
will form a certain plane angle $\theta$ with the $z$ axis. Changing the value 
of the angle $\varphi$ from zero to $\Phi$, we get a set of plane angles 
$\theta(\varphi)$ that fill the entire solid angle under consideration.
This process is analogous to the process of formation of a flat 
two-dimensional region by a set of straight line segments or a three-dimensional 
object by a set of two-dimensional flat figures. This means that the 
solid angle is a derived quantity in the SI formed by plane angles, just as 
the area is a derived quantity formed by lengths. It follows thence that the 
coherent unit of the solid angle in the SI is rad$^2$. 

Further, the connection between a unit of plane angle and the steradian is 
described. The whole set of directions of the rays defining the conical surface 
of the solid angle will be determined by the function $\theta(\varphi)$. 
The value of the solid angle $\omega$ is obtained by integrating the element 
of the solid angle $d\omega=\sin\theta d\theta d\varphi$ over the region 
$0\le\varphi\le\Phi$, $0\le\theta\le\theta(\varphi)$
\begin{equation}
\label{omega}%
\omega = \int_{0}^{\Phi}\int_{0}^{\theta(\varphi)}\sin\theta d\theta d\varphi.
\end{equation}                                      
Here, as before, $\Phi$ is the full plane angle. Integration over the $\theta$ 
is easily performed, giving expression
\begin{equation}
\label{omega-teta/fi}%
\omega = \frac{\Phi}{2\pi}\int_{0}^{\Phi}[1-\cos\theta(\varphi)]d\varphi.
\end{equation}                                      
The coefficient $\Phi/2\pi$ appears when integrating the trigonometric function, 
as it was noted in Section 2.

Using this expression, we can find the value of the total solid angle. Let our 
function $\theta(\varphi)$ be a constant $\theta$. The corresponding solid 
angle is a region of space inside a circular cone. The integral in the 
\eqref{omega-teta/fi} gives the following value of the solid angle 
\begin{equation}
\label{omega-teta}%
\omega = \frac{\Phi^2}{2\pi}(1-\cos\theta).
\end{equation}                                      

At $\theta=0$ (the conical surface degenerates into the axis O$z$), the 
corresponding solid angle will be zero. The total solid angle $\Omega$ is 
obtained at the maximum value of the angle $\theta = \Phi/2$, for which 
$\cos(\Phi/2)=-1$. The expression \eqref{omega-teta} gives the following value 
for $\Omega$
\begin{equation}
\label{Omega}%
\Omega = \Phi^2/\pi.
\end{equation}                                      

Comparing this expression with the formula \eqref{steradian}, we obtain
\begin{equation}
\label{sr-fi}%
1\, \mbox{sr} = \frac{\Phi^2}{4\pi^2}.
\end{equation}                                      
If a radian measure is used for a plane angle, this expression takes the form
\begin{equation}
\label{sr-rad}%
1\, \mbox{sr} = 1\, \mbox{rad}^2.
\end{equation}                                     
This means that in the current SI, the steradian is a coherent derived unit of 
the radian, but not a meter.

The SI definition of candela adopted as far back as 1948  was replaced in 1979 
with a new definition, in which the unit of solid angle, the steradian, was 
included. At that time, the steradian had a nonzero dimension and belonged to 
the class of additional units, the status of which caused discussions. In 1995, 
the steradian was declared as a coherent dimensionless derived  unit of solid angle. 
In spite of the transformation of the steradian, the candela remained in the list 
of the base SI units.

In the draft of 9th edition of the SI Brochure, six base units are defined 
through a set of defining constants \cite[P. 13-17]{SI-9}, but the definition 
of the candela contains, in addition to the defining constants, also the derived 
unit -- the steradian
\begin{equation}
\label{candela}%
1\, \mbox{cd} = 2,614830\cdot10^{10}\cdot (\Delta\nu_{Cs})^2h
K_{\mbox{\footnotesize cd}} \,\mbox{sr}^{-1},                         
\end{equation}
where $h$, $\Delta\nu_{Cs}$, and $K_{\mbox{cd}}$ are defining constants for the 
base SI units, having the fixed values. In that draft the steradian is assumed to 
be equal to a dimensionless number one for that expression, 
and an expression that does not contain the steradian is obtained for candela
\begin{equation}
\label{candela-si}%
1\, \mbox{cd} = 2,614830\cdot10^{10}\cdot (\Delta\nu_{Cs})^2h
K_{\mbox{\footnotesize cd}} \qquad \mbox{(in the New SI)}.    
\end{equation}                                            

These results produce a lot of questions.
Does this candela correspond to its 1979 definition? Can we continue to consider 
the candela as a base and coherent SI unit? What about redefinitions of 
the base units, which include the candela defined in this way? Is the current list 
of base SI units complete? These questions require the thorough investigation.

\section{Conclusion}

The findings of the studies carried out in this paper are as follows.


\begin{enumerate}

\item  The reasoning given in CIPM Recommendation 1 (1980) for 
transferring the plane and solid angles into the class of dimensionless derived 
quantities is unfounded.

\item   The plane and solid angles are quantities of different kinds, having 
different geometric dimensions. And, consequently, they have different units of 
measurement, that do not coincide with the dimensionless number one.

\item   The plane angle is a quantity independent of other SI quantities. It
should be included, most likely, into the base quantities of the SI.

\item      The units of the radian and the steradian can be determined by fixing 
the exact values of the total plane $\Phi$ and solid $\Omega$ angles:

\begin{itemize}

\item      the radian is defined on the condition that the total plane angle 
is equal to $\Phi = 2\pi$ rad,

\item      the steradian is defined on the condition that the total solid 
angle is equal to $\Omega = 4\pi$ sr.

\end{itemize}

\item     The solid angle in the SI is a derived quantity of the plane angle,
not the lengt. Its coherent unit is the steradian, equal to the squared radian.

\item     The dependence of the candela definition on steradian produces 
a lot of questions about its status.

\end{enumerate}


\begin{thebibliography}{9}

\bibitem{res12-cgpm11}
Resolution 12 of the 11th CGPM (1960). \\
https://www.bipm.org/en/CGPM/db/11/12/ .

\bibitem{res8-cgpm20}
Resolution 8 of the 20th CGPM (1995).
https://www.bipm.org/en/CGPM/db/20/8/.

\bibitem{SI-8}
The International System of Units (SI-8). 8th edition, 2006.\\
https://www.bipm.org/utils/common/pdf/si\_brochure\_8\_en.pdf .

\bibitem{recom-u1-ccu_80}
7$^{\rm e}$ Session Comit\'e Consultatif des Unit\'es. 1980. Recommandation U 1. 
Unit\'es suppl\'ementaires radian et st\'eradian.\\
https://www.bipm.org/utils/common/pdf/CC/CCU/CCU7.pdf .

\bibitem{recom1-cipm80}
CIPM, 1980: Recommendation 1. \quad   https://www.bipm.org/en/CIPM/db/1980/1/ .

\bibitem{SI-9}
The International System of Units (SI) 9th ed. 2019 (Draft).\\
https://www.bipm.org/utils/en/pdf/si-revised-brochure/Draft-SI-Brochure-2018.pdf

\bibitem{Wittmann}
H. Wittmann. A New Approach to the Plane Angle. Metrologia, 1988, 
V. 25, No 4, P. 193-203.

\bibitem{Maxwell}
J.C. Maxwell. Treatise on Electricity and Magnetizm. Oxford, 
Oxford University Press, 1873.

\bibitem{mathencycl-4} %
Mathematical encyclopedia, Vol. 4. Moscow, Publishing house Sovetskaya 
encyclopedia, 1984, (in Russian).

\bibitem{mathencycl-5} %
Mathematical encyclopedia, Vol. 5. Moscow, Publishing house Sovetskaya 
encyclopedia, 1985, P. 326, (in Russian).


\end{thebibliography}
\end{document}